\documentclass[12pt,preprint]{aastex}

\newcommand       \Angstrom     {\,{\rm \AA}}

\newcommand   \pc       {\,{\rm pc}}

\newcommand       \simgt        {\gtrsim}

\newcommand       \mum          {\,{\rm \mu m}}

\newcommand       \simali       {\sim\,}
\shorttitle{Dust Extinction of GRB Host Galaxies}

\begin{document}
\title{
Dust Extinction of Gamma-ray Burst Host Galaxies:
Identification of Two Classes?
     }
\author{S.L. Chen\altaffilmark{1},
        Aigen Li\altaffilmark{1,2},
        and D.M. Wei\altaffilmark{1,3}}

\altaffiltext{1}{Purple Mountain Observatory, Chinese Academy of
                 Sciences, Nanjing 210008, China; {\sf fugue@pmo.ac.cn}}
\altaffiltext{2}{Department of Physics and Astronomy, University of
                 Missouri, Columbia, MO 65211; {\sf lia@missouri.edu}}
\altaffiltext{3}{Joint Center for Particle Nuclear Physics and
                 Cosmology of Purple Mountain Observatory --
                 Nanjing University, Nanjing 210008, China;
                 {\sf dmwei@pmo.ac.cn}}

\begin{abstract}
Dust in the host galaxies of gamma-ray bursts (GRBs)
dims and reddens their afterglow spectra.
Knowledge of the nature of this dust
is crucial for correcting for extinction,
providing clues to the nature of GRB progenitors,
and probing the interstellar medium of high-redshift galaxies
as well as the nature of cosmic dust when the universe was
much younger and galaxies were much less evolved.
The dust and extinction properties of GRB host galaxies
are still poorly known. Unlike previous work, we derive
in this {\it Letter} the extinction curves for 10 GRB host
galaxies without a priori assumption of any specific
extinction types (such as that of the Milky Way,
or the Small/Large Magellanic Clouds).
It is found that there appears to exist two different types
of extinction curves: one is relatively flat and gray,
the other displays a steeper dependence on inverse
wavelength, closely resembling that of the Milky Way but
with the 2175$\Angstrom$ feature removed.
\end{abstract}


\section{Introduction}
The existence of dust extinction towards gamma-ray bursts (GRBs)
and their afterglows has been well established
{\it observationally} through
(1) ``dark bursts'' -- nearly $\simali$60\% of the X-ray afterglows
have no optical counterparts -- the undetected optical afterglows
may have been extinguished by dust in the host galaxy
(see Lazzati et al.\ 2002 and references therein);
(2) reddening -- the optical/near-infrared (IR) spectral energy
distributions (SEDs) of afterglows deviate from that expected
from standard models,\footnote{%
  As any host-galaxy extinction would result
  in a steepening of the intrinsic spectral slope,
  a steeper (observed) slope (than all intrinsic spectral
  slopes) betrays the presence of host galaxy dust.
  }
indicative of dust reddening
(see Kann et al.\ 2006 and references therein);
(3) the reduced Balmer line ratios
(as a consequence of dust extinction)
in the spectra of some GRB host galaxies
(e.g. see Djorgovski et al.\ 1998)
compared with the expected ratios for
the standard Case B recombination (Osterbrock \& Ferland 2006);
(4) the depletion of dust-forming heavy elements
such as Si and Fe in some host galaxies
(e.g. see Savaglio et al.\ 2003);
and {\it theoretically} (5) through the association of
(at least long-duration [$\simgt$\,2\,s]) GRBs
with massive stars and star-forming regions
embedded in dense clouds of dust and gas (Paczy\'nski 1998).

A precise knowledge of the nature (e.g. size, composition)
of the dust in GRB host galaxies is very useful for
(1) correcting for the extinction of afterglows
from X-ray to near-IR wavelengths;
(2) constraining the nature of the GRB progenitors
(i.e. collapsing massive stars or merging neutron binaries);
and (3) understanding the interstellar medium (ISM) of
high-redshift galaxies and the cosmic star formation history
(e.g. see Ramirez-Ruiz et al.\ 2002).

However, our knowledge of the dust in GRB host galaxies
is very limited. Previous studies all are limited to a qualitative
and rather speculative analysis of the extinction curves derived by
fitting the observed optical/near-IR SEDs of GRB afterglows with a
power-law (approximating their intrinsic spectra) reddened by an
extinction law of known types (mostly either that of the Milky Way
[MW], the Small Magellanic Cloud [SMC], or that of the Large
Magellanic Cloud [LMC]) adopted as a priori (e.g. see Stratta et
al.\ 2004, Kann et al.\ 2006).

In this {\it Letter}, by assuming that the GRB afterglows
have a power-law intrinsic spectrum as expected from
the standard fireball model, we obtain the extinction curves
for 10 GRB host galaxies without a priori assumption of
the extinction law. The size and composition of the dust
are modeled quantitatively in terms of the silicate-graphite
interstellar grain model.

\section{Method}
In the standard fireball model, the temporal ($t$)
and frequency ($\nu$) dependence of GRB afterglows
can be well described by
$F_\nu(t)\propto t^{-\alpha} \nu^{-\beta}$,
where both $\alpha$ and $\beta$ are related with
the electron energy distribution index $p$
($dn\propto E^{-p}\,dE$ for the energy distribution
of the shock-heated electrons; see Sari et al.\ 1998).
Therefore, with the decay index $\alpha$ determined,
$p$ as well as the intrinsic spectral index $\beta$
can be well constrained.

In addition to the extinction of the host galaxy,
GRB afterglows are also subject to the extinction
of our own MW galaxy. We correct for the latter
using the reddening maps of
Schlegel et al.\ (1998).
The host's spectrum is redshifted to longer wavelengths
where any errors in the Schlegel et al.\ map are not important.

Let $F_{\lambda}^{0}$ be the intrinsic spectrum. With the Galactic
reddening corrected, the observed wavelength ($\lambda$)-dependent
spectrum is
$F_\lambda\,=\,F_{\lambda}^{0}\,\exp\left(-A_\lambda\right)$, where
$A_\lambda$ is the extinction from the host galaxy. Setting $V$ band
as the zero point, we obtain
$F_\lambda/F_V\,=\,\left(\lambda_V/\lambda\right)^{2-\beta_0}
\exp\left(A_V-A_\lambda\right)$, where $\lambda$ and
$\lambda_V$\,$\equiv$\,5500$\Angstrom$ are both measured in the rest
frame. With $\beta_{0}$ estimated from the decay index $\alpha$, we
can obtain $A_\lambda - A_V$ from $F_\lambda/F_V$. To determine
$A_V$, we fit a simple function $A_\lambda - A_V=
a/\left[1+\left(\lambda/\lambda_B\right)^2\right]+b$ for $\lambda >
\lambda_V$ and then extrapolate this function to $\lambda\rightarrow
\infty$ to get $A_V$ (=$-b$) since  $A_\lambda\rightarrow 0$ as
$\lambda\rightarrow \infty$.

The standard silicate-graphite interstellar dust model,
consisting of a mixture of spherical silicate and graphite grains,
is shown successful in reproducing the extinction and IR
emission of the MW galaxy, SMC and LMC
(Weingartner \& Draine 2001; Li \& Draine 2001, 2002).
We will apply this grain-mixture to model the extinction curves
$A_\lambda/A_V$ determined for the GRB host galaxies, but with
a {\it simpler} functional formula
for the dust size distribution:
$dn\sim a^{-\eta}\,\exp(-a/a_c)\,da$
for both grain types, where $a$ is the grain radius,
ranging from 0.05 to 2.5$\mum$,
$\eta$ is the power-law index,
$a_c$ is the cut-off size, and $dn$ is the number of grains
in the size interval [$a$,\,$a+da$].
The mass fraction of graphite dust is $f_{\rm gra}$
[for silicates it is $(1-f_{\rm gra})$].
In Figure 1 we show that the MW, SMC and LMC
extinction curves are well fit by the silicate-graphite
model with this simple dust size distribution function.

\section{Data}
To construct an extinction curve, one requires simultaneous
multi-band photometry. We carefully collected such data from
literature for ten bursts taken when their decays were in a steady
power-law state(e.g. see Panaitescu \& Kumar 2001, Fan \& Piran 2006
for very detailed analysis). The optical and near-IR UBVRIJHK fluxes
$F_\nu$ of these bursts are tabulated in Table 1. Also tabulated is
$\beta_0$, the intrinsic spectrum index calculated from fitting the
decay index based on the standard model of afterglow (see \S2).

To compare the extinction properties in the rest frame
of each burst, in the observer frame we need to calculate
$A_\lambda - A_{[V(1+z)]}$.
The results are shown in Table 2.
We correct the wavelength with a factor $(1+z)$ to
get $A_\lambda - A_V$ in the rest frame.
Also shown in Table 2 are $A_V$, obtained by extrapolating
$A_\lambda - A_V$ to $\lambda \rightarrow \infty$
(see \S2).

\section{Extinction Curves: Two Different Types?}
The extinction curves (normalized to the $V$ band)
derived for these 10 host galaxies are shown in
Figures 2 and 3. A common feature for these curves
is the absence of the 2175$\Angstrom$ bump,
which is the strongest absorption band
in the MW extinction (Li 2005).
They appear to fall into two categories:
one is relatively gray, showing a much flatter increase
with inverse wavelength $\lambda^{-1}$
(we call it ``Type-I''; see Fig.\,2),
the other displays a much steeper,
almost linear increase with $\lambda^{-1}$
(we call it ``Type-II''; see Fig.\,3).

In view of its possible practical use (e.g. in correcting for
extinction of afterglows), following Fitzpatrick \& Massa (1990),
we take an analytical fit to these 2 extinction types:
\begin{equation}
A_\lambda/A_V =
\left\{\begin{array}{ll}
c_1 + c_2\,x + c_3\,D(x;\gamma,x_0) + c_4 F(x), & x\ge 1\mum^{-1}, \\
k\,x^{1.84}, & x\,<\,1\mum^{-1},\\
\end{array} \right .
\end{equation}
where $x\equiv \lambda^{-1}$,
$D(x;\gamma,x_0)\,=\,x^2/\left[\left(x^2-x^2_0\right)^2
+ x^2\gamma^2\right]$, and
\begin{equation}
F(x) =
\left\{\begin{array}{ll}
0.5392\,\left(x-5.9\right)^2 + 0.05644\,\left(x-5.9\right)^3,
& x\ge 5.9\mum^{-1},\\
0, & x\,<\,5.9\mum^{-1}.\\
\end{array} \right .
\end{equation}
In this fit, $c_1$ and $c_2$ determine a linear ``background''
term; $c_3$ determines the strength of the 2175$\Angstrom$
feature which is represented by the ``Drude profile'' term
$D(x;\gamma,x_0)$ -- the theoretical profile
for a classic damped harmonic oscillator
(Bohren \& Huffman 1983);
$\gamma$ and $x_0$ are respectively
the FWHM and peak position of the Drude profile;\footnote{%
  In the Galactic ISM, the strength and width of
  the 2175$\Angstrom$ extinction bump vary with
  environment while its peak position is quite invariant
  (see Li 2005).
  We therefore fix $x_0$ at 4.6$\mum^{-1}$.
  }
$c_4$ determines the far-UV curvature term represented
by $F(x)$.
The resulting analytical fits are plotted in Figures 2 and 3;
and the fitted parameters $c_1$, $c_2$, $c_3$, $c_4$,
and $\gamma$ are tabulated in Table 3.

To quantify the dust properties, we fit both extinction types
in terms of the silicate-graphite dust model.
The model fits and parameters are respectively
shown in Figures 2,3 and in Table 4.
The model for the flat ``Type-I'' extinction curve
is dominated by {\it large} silicate grains
(with $f_{\rm gra} \approx 0$ and a flat size distribution).
In contrast, the model for the steep ``Type-II'' extinction
has a much steeper size distribution, indicating the richness
of {\it small} dust in the host galaxies
with a ``Type-II'' extinction curve.

\section{Discussion}
Previous efforts in deriving the dust extinction of GRB host
galaxies all assume an initial power-law for the GRB intrinsic
spectrum (as expected from the standard fireball model), and an
extinction curve of known types such as that of MW, SMC, LMC,
the Calzetti et al.\ (1994) law
suitable for local starburst galaxies,
and the Maiolino et al.\ (2001) law
suggested for the dust in the circumnuclear
region of AGNs (e.g. see Stratta et al.\ 2004 and Kann et al.\ 2006
for recent examples).
Our approach, without the need for an initial assumption
of specified extinction types, is more favourable because
of the lack of a priori knowledge of the extinction in the GRB
hosts. The fact that neither of the two extinction types derived in
this work resembles that of MW, SMC and LMC (see Figs.\,2,3)
challenges the initial assumption of specific extinction types.

In consistent with previous studies, this work
also finds no evidence for the 2175$\Angstrom$
extinction bump.\footnote{%
  The only exception is GRB 991216.
  Its afterglow spectrum shows a depression
  in flux in between 4000$\Angstrom$ and 5500$\Angstrom$,
  suggesting the presence of a {\it red}
  2175$\Angstrom$-type extinction bump
  (at $\simali$2360\AA) in the host of GRB 991216
  (Vreeswijk et al.\ 2006).
  Similar red 2175$\Angstrom$ bumps have been previously
  seen in a few UV-strong, hydrogen-poor stars
  in the MW galaxy (see Li 2005 and references therein).
  }
The carrier of this bump remains
unidentified over 40 years after its first detection.
It is generally believed to be caused by aromatic
carbonaceous (graphitic) materials,
likely a cosmic mixture of polycyclic aromatic
hydrocarbon (PAH) molecules
(see Li 2005 and references therein).
The nondetection of the 2175$\Angstrom$ extinction bump
in the GRB hosts could be explained in terms of
(1) the depletion of its carrier (e.g. PAHs)
by condensation onto the ice mantles coated
on the dust in the dense star-forming regions
in which GRBs are embedded, and
(2) the destruction of its carrier by UV/X-ray
radiation in the immediate vicinities (up to
$\simali$20\pc) of bursts.

Our Type-I extinction curve (see Fig.\,2) is more
gray (i.e. with a much weaker dependence on wavelength)
than that of MW, SMC, and LMC.
Gray dust has been invoked by a number of authors
(e.g. Savaglio et al.\ 2003, Savaglio \& Fall 2004;
Stratta et al.\ 2004, 2005)
to account for the small ratios of extinction
and/or reddening (derived from fitting the afterglow SEDs)
to H column densities (determined from X-ray or
Ly$\alpha$ absorption) of GRB hosts, which are
usually smaller than that of MW
by a factor of $\simali$10--100
(Galama \& Wijers 2001, {\v S}imon et al.\ 2001,
Hjorth et al.\ 2003, Vreeswijk et al.\ 2004).\footnote{%
  Alternatively, these results can be explained by
  an intrinsicaly low dust content --
  a low metallicity and/or a low level of depletion
  of the dust-forming heavy elements in the burst
  environment (Fynbo et al.\ 2003, Hjorth et al.\ 2003,
  Vreeswijk et al.\ 2004, Fruchter et al.\ 2006).
  However, Savaglio et al.\ (2003) found that in
  the host galaxies of 3 GRBs
  both the column densities of metals (as indicated by Zn)
  and the depletion of heavy elements (such as Fe, Si and Cr)
  are large, indicating a large dust content.
  }
Gray dust could be created by
(1) the preferential destruction of small grains
by the intense UV and X-ray radiation from the GRB
(Waxman \& Draine 2000, Fruchter et al.\ 2001,
Perna et al.\ 2003) in the immediate GRB
environment, say, within $\simali$10--20\,pc around the burster
(Hjorth et al.\ 2003, Savaglio et al.\ 2003, Stratta et al.\ 2005),
and (2) the growth of dust through coagulation
(which also leads to the depletion of small grains)
in the high-density environments such as the cores of
star-forming regions harboring GRB events
(Maiolino et al. 2001, Stratta et al.\ 2004).
Both mechanisms naturally lead to a dust size distribution
skewed toward large grains, producing an extinction curve
weakly dependent on wavelength.

Nongray extinction has also been reported
(e.g. for the host of GRB 010222 by Galama et al.\ 2003).
Our nongray, Type-II extinction curve (see Fig.\,3)
shows a steep increase with $\lambda^{-1}$,
indicating that small grains are more abundant
in the GRB host galaxies with a Type-II
extinction than in those with a Type-I extinction.
This is probably caused by a less complete destruction of
small grains (e.g. associated with less energetic bursts
of less intense UV/X-ray radiation), and/or a slower
coagulational growth of dust (e.g. associated with less dense
environments or metal-poorer galaxies).
Since our sample is small, we do not want to
overinterpret its implications.
A more systematic investigation of the extinction
and dust properties of a large sample of
GRB host galaxies is in progress.

Finally, if one really wants to make a priori assumption
of the extinction spectral shape when deriving the GRB host
galaxy dust extinction, we suggest the adoption of
the Fitzpatrick \& Massa (1990) formulae instead of
that of MW, SMC or LMC. By first visually inspecting
the GRB spectrum, one can set $c_3=0$ (in eq.[1])
if there is no sign of the 2175$\Angstrom$ extinction
bump or curvature on large wavelength intervals.

\acknowledgments
We thank Y.Z. Fan, D.A. Kann, G. Stratta, and the anonymous
referee for helpful comments. S.L.C. and D.M.W. are supported
by the NSFC grants 10225314 and 10233010,
and the National 973 Project NKBRSF G19990754.
A.L. is supported by the University of Missouri
Summer Research Fellowship, the University of Missouri
Research Board, and NASA/Spitzer Theory Programs.

\clearpage

\begin{figure}
\plotone{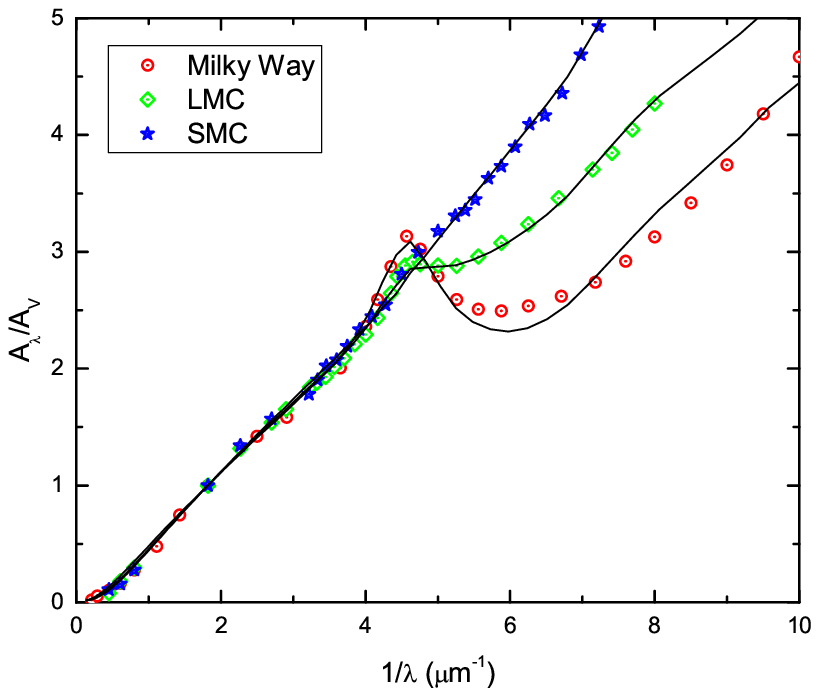}
\caption{
        \label{fig:mwsmclmc}
        Fitting the extinction curves
        of the Milky Way galaxy, the SMC and LMC
        by the silicate-graphite model with
        a simple dust size distribution
        $dn/da\propto a^{-\eta}\,\exp(-a/a_c)$.
        Points -- observational data;
        solid lines -- model fits.
        }
\end{figure}

\clearpage

\begin{figure}
\plotone{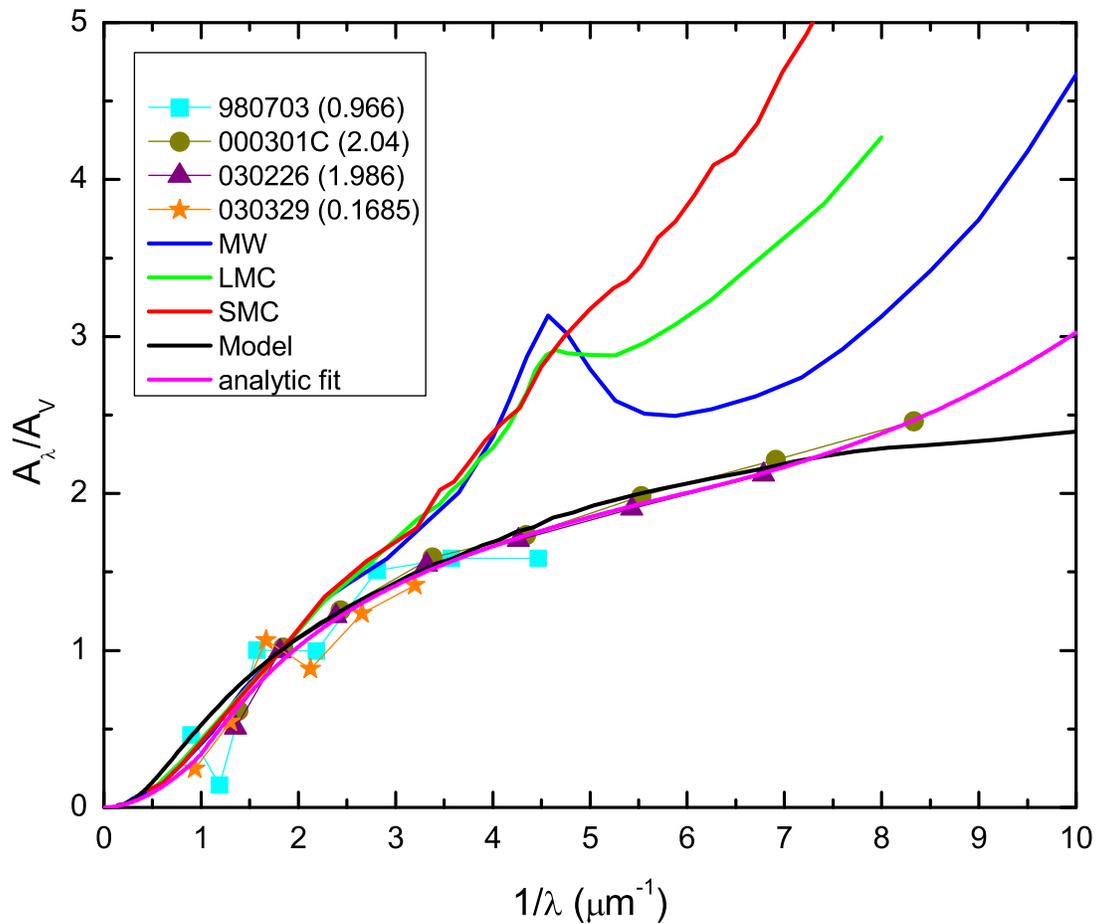}
\caption{
        \label{fig:typeI}
        Flat, ``Type-I'' extinction curves
        (normalized to the $V$ band) for 4 GRB host galaxies.
        Both axes are in the rest-frame of the GRB host.
        Also plotted are the model (black) and analytical (magenta)
        fits to the observed extinction curves.
        The MW, SMC and LMC extinction curves
        are also shown for comparison.
        }
\end{figure}

\clearpage

\begin{figure}
\plotone{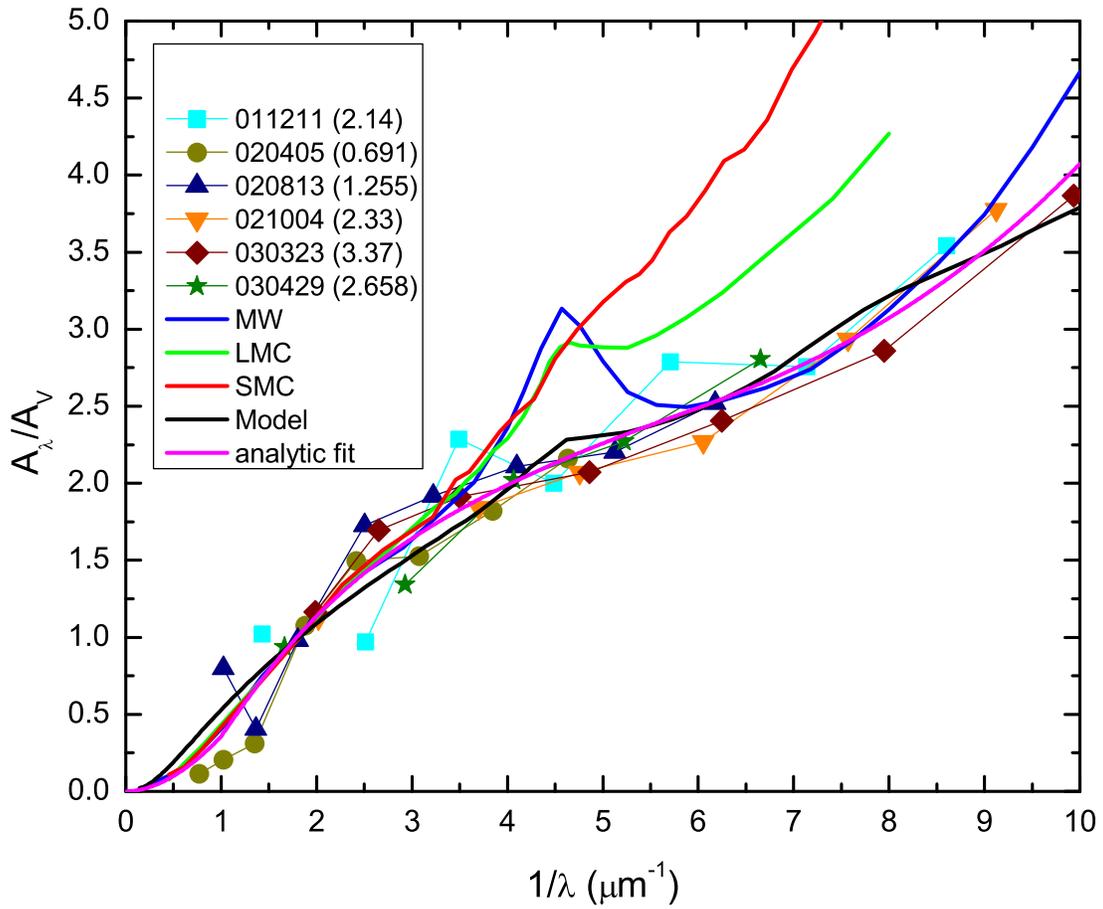}
\caption{
        \label{fig:typeII}
        Same as Figure 2 but for the relatively steep,
        ``Type-II'' extinction curves derived for 6 GRB
        host galaxies.
        }
\end{figure}

\clearpage

\begin{table}
\caption[]{Optical to near-IR fluxes of 10 GRB afterglows\label{tab:tbl1}}
{\footnotesize
\begin{tabular}{cccccccccccccc}
\tableline\tableline GRB & $z$ &  $\beta_0$ & $A_V^{\rm MW}$
&\multicolumn{8}{c}{$F_\nu$ ($\mu Jy$)} & $t$ & References\\
\cline{5-12}
  &    &  & (mag) & $U$ & $B$ & $V$ & $R$ & $I$  &$J$   &$H$  & $K$ & (days)\\
\cline{5-12} \tableline
980703   & 0.966  & 1.07 &0.192 & \nodata  &2.7  &3.4  &4.6  &7.6   &10.8  &21.8  &25.5  & 5   & 1 \\
000301C  & 2.0404 & 0.56 &0.308 &11.6  &14.1 &17.3 &21.6 &26.1  &35.3  &44.9  &60.6  &3.1  & 2 \\
011211   & 2.14   & 0.63 &0.142 &11.4  &13.5 &15.5 &19.1 &21.9  &29.5  &\nodata &42.0  &0.57 & 3 \\
020405   & 0.691  & 0.5  &0.180 &5.4   &7.2  &9.5  &10.9 &15.7  &28.6  &34.9  &42.5  &1.98 & 4 \\
020813   & 1.255  & 0.52 &0.370 &6.1   &7.7  &9.0  &11.1 &13.7  &22.4  &33.2  &32.5  &1.93 & 5 \\
021004   & 2.33   & 0.45 &0.198 &4.3   &7.2  &11.1 &13.7 &17.2  &\nodata &32.6  &\nodata &5.5  & 6 \\
030226   & 1.986  & 0.33 &0.062 &\nodata &17.3 &20.4 &23.9 &27.6  &35.1  &42.1  &56.2  &0.8  & 7 \\
030323   & 3.3718 & 0.28 &0.163 &\nodata &2.0  &5.2  &7.6  &10.8  &16.5  &21.0  &37.7  &2.4  & 8 \\
030329   & 0.1685 & 0.48 &0.083 &1200   &1400 &1770 &1860 &2530  &3300  &\nodata &\nodata &0.7  & 9 \\
030429   & 2.658  & 0.63 &0.204 &\nodata &\nodata &11.2 &16.3 &21.1  &34.3  &\nodata &57.8  &0.548& 10 \\
\tableline
\end{tabular}
\tablecomments{Data are all taken from the cited references.
The Galactic visual extinction $A_V^{\rm MW}$ is derived from
the reddening maps of Schlegel et al.\ (1998).
The UBVRIJHK fluxes are measured in the observer frame;
$z$ is the redshift of the burst; $t$ is the time when the data was
taken since the burst; and $\beta_0$ is the intrinsic spectral index
derived from the standard afterglow model (see \S2).} \tablerefs{(1)
Frail et al.\ 2003; (2) Jensen et al.\ 2001; (3) Jakobsson et al.\
2003; (4) Stratta et al.\ 2005; (5) Covino et al.\ 2003; (6) Holland
et al.\ 2003; (7) Klose et al.\ 2004; (8) Vreeswijk et al.\ 2004;
(9) Bloom et al.\ 2004; (10) Jakobsson et al.\ 2004.} }
\end{table}

\clearpage

\begin{table}
\caption[]{Extinction in the rest frame of each burst
calculated from Table 1. $A_V$ is obtained by extrapolating
an analytical fit of $A_\lambda-A_{\left[V(1+z)\right]}$ to
$\lambda \rightarrow \infty$ (see \S2).\label{tab:tbl2}}
{\footnotesize
\begin{tabular}{ccccccccccccc}
\tableline\tableline
GRB & $z$ & $\beta_0$ & $A_V$ &\multicolumn{8}{c}{$A_\lambda-A_{[V(1+z)]}$}\\
\cline{5-12}
  & &     &           & $U$  & $B$  & $V$  & $R$  & $I$  &$J$   &$H$   & $K$\\
\cline{4-11} \tableline
980703   & 0.966   & 1.07   &0.48    &\nodata  &0.29  &0.29  &0.24  &0     &0     &-0.41 &-0.26   \\
000301C  & 2.0404  & 0.56   &0.35    &0.51  &0.43  &0.34  &0.26  &0.21  &0.1   &0.01  &-0.13    \\
011211   &2.14     &0.63    &0.07    &0.18  &0.12  &0.13  &0.07  &0.09  &-0.002& \nodata &0.002    \\
020405   &0.691    & 0.5    &0.57    &0.66  &0.47  &0.30  &0.28  &0.04  &-0.39 &-0.45 &-0.51    \\
020813   & 1.255   & 0.52   &0.43    &0.65  &0.52  &0.48  &0.39  &0.31  &-0.01 &-0.26 &-0.09     \\
021004   &2.33     &0.45    &0.51    &1.42  &0.99  &0.65  &0.55  &0.43  &\nodata &0.06  &\nodata    \\
030226   &1.986    &0.33    &0.40    &\nodata &0.45  &0.36  &0.28  &0.22  &0.09  &-0.002&-0.20     \\
030323   & 3.3718  &0.28    &0.92    &\nodata &2.64  &1.71  &1.29  &0.99  &0.84  &0.64  &0.15      \\
030329   &0.1685   &0.5     &0.36    &0.14  &0.08  &-0.05 &0.03  &-0.16 &-0.26 &\nodata  &\nodata         \\
030429   &2.658    &0.63    &0.41    &\nodata &\nodata &0.74  &0.52  &0.42  &0.14  &\nodata  &-0.03      \\
\tableline
\end{tabular}
}
\end{table}

\clearpage

\begin{table}
\caption[]{Parameters for fitting the extinction curves of
           GRB host galaxies with the Fitzpatrick \& Massa
           (1990) formulae. Note that $x_0$ is fixed at
           $4.6\mum^{-1}$ and is not a free parameter
           (see Eq.[1]).\label{tab:tbl3}}
\begin{tabular}{ccccccc}
\tableline\tableline
 & $c_1$ & $c_2$ & $c_3$ & $c_4$ & $\gamma$ ($\mum^{-1}$) & $k$ \\
\cline{1-7}
Type-I     & -0.342 & 0.181  & 331.86  &0.040 &16.040 &0.340  \\
Type-II    & -0.343 & 0.263  & 268.34  &0.061 &14.431 &0.358   \\
\tableline
\end{tabular}
\end{table}

\clearpage

\begin{table}
\caption[]{Parameters for modeling the 2-type extinction curves
           of GRB host galaxies (as well as that of MW, SMC and LMC)
           with the silicate-graphite dust model.\label{tab:tbl4}}
\begin{tabular}{ccccc}
\tableline\tableline
Galaxy & $\eta$ & $a_c$ ($\mu$m) & $f_{\rm gra}$
       & $\chi^2/{\rm dof}$\tablenotemark{a}\\
\cline{1-5}
MW         & 3.14        & 0.17     & 0.70       & 0.30     \\
LMC        & 3.16        & 0.20     & 0.29       & 0.04      \\
SMC        & 3.33        & 0.26     & 0.02       & 0.06     \\
Type-I     & 2.61        & 0.21     & 0          & 0.63    \\
Type-II    & 3.09        & 0.29     & 0.22       & 1.97     \\
\tableline
\end{tabular}
\tablenotetext{a}{\footnotesize $\chi^2 \equiv \sum_{\rm all\,\lambda}\sum_{\rm all\,GRBs}\left[ \left(A_\lambda/A_V\right)_{\rm mod} - \left(A_\lambda/A_V\right)_{\rm obs}\right]^2/\sigma^2$ is obtained by summing up
all wavebands and all GRBs, where $\sigma$ is the uncertainty
for a given GRB at a given band.}
\end{table}
\end{document}